\documentclass[lettersize,journal]{IEEEtran}
\usepackage{array}
\usepackage[caption=false,font=normalsize,labelfont=sf,textfont=sf]{subfig}
\usepackage{textcomp}
\usepackage{stfloats}
\usepackage{url}
\usepackage{verbatim}
\usepackage{amsmath,amsfonts,amssymb}
\usepackage{graphicx}
\usepackage{setspace}
\usepackage[subfigure]{tocloft}
\usepackage{subfig}
\usepackage{algorithm}
\usepackage{algpseudocode}
\usepackage{multirow}
\usepackage{multicol}
\usepackage{bbding}
\usepackage{pifont}
\usepackage{amsmath}
\usepackage{xcolor} 

\usepackage{bm} 

\usepackage{etoolbox}
\makeatletter
\patchcmd{\@makecaption}
  {\scshape}
  {}
  {}
  {}
\makeatletter

\usepackage[utf8]{inputenc} 
\usepackage[T1]{fontenc}    
\usepackage{hyperref}       
\usepackage{url}            
\usepackage{booktabs}       
\usepackage{amsfonts}       
\usepackage{nicefrac}       
\usepackage{microtype}      

\DeclareMathOperator*{\argmax}{arg\,max}

\DeclareMathOperator{\N}{\mathcal{N}}
\DeclareMathOperator{\I}{\mathbf{I}}
\DeclareMathOperator{\x}{\mathbf{x}}

\def\BibTeX{{\rm B\kern-.05em{\sc i\kern-.025em b}\kern-.08em
    T\kern-.1667em\lower.7ex\hbox{E}\kern-.125emX}}
\usepackage{balance}

\begin{document}

\title{Pi-fusion: Physics-informed diffusion model for learning fluid dynamics}

\author{{Jing Qiu, Jiancheng Huang, Xiangdong Zhang, Zeng Lin, Minglei Pan, Zengding Liu, Fen Miao$^{*}$,~\IEEEmembership{Member,~IEEE}}


\thanks{Manuscript created January, 2024. This work was supported in part by the National Natural Science Foundation
of China under Grant U2241210, and in part by the Basic Research Project of Shenzhen under Grants JCYJ20220818101216034 and JCYJ20210324101206017(corresponding author: Fen Miao, email: fen.miao@siat.ac.cn).

Jing Qiu, Jiancheng Huang, Minglei Pan and Zengding Liu are with Shenzhen Institute of Advanced Technology, Chinese Academy of Sciences, Shenzhen 518055, China and also with the University of Chinese Academy of Sciences, Beijing 101408, China.

Xiangdong Zhang is with University of Macau, Macau, China and also Shenzhen Institute of Advanced Technology, Chinese Academy of Sciences, Shenzhen 518055, China.

Zeng Lin is with Shenzhen Institute of Advanced Technology, Chinese Academy of Sciences, Shenzhen 518055, China.

Fen Miao is with Shenzhen Institute of Advanced Technology, Chinese Academy of Sciences and also with University of Electronic Science and Technology of
China, Chengdu 611731, China}}
\maketitle

\begin{abstract}
Physics-informed deep learning has been developed as a novel paradigm for learning physical dynamics recently. While general physics-informed deep learning methods have shown early promise in learning fluid dynamics, they are difficult to generalize in arbitrary time instants in real-world scenario, where the fluid motion can be considered as a time-variant trajectory involved large-scale particles. Inspired by the advantage of diffusion model in learning the distribution of data, we first propose Pi-fusion, a physics-informed diffusion model for predicting the temporal evolution of velocity and pressure field in fluid dynamics. Physics-informed guidance sampling is proposed in the inference procedure of Pi-fusion to improve the accuracy and interpretability of learning fluid dynamics. Furthermore, we introduce a training strategy based on reciprocal learning to learn the quasiperiodical pattern of fluid motion and thus improve the generalizability of the model. The proposed approach are then evaluated on both synthetic and real-world dataset, by comparing it with state-of-the-art physics-informed deep learning methods. Experimental results show that the proposed approach significantly outperforms existing methods for predicting temporal evolution of velocity and pressure field, confirming its strong generalization by drawing probabilistic inference of forward process and physics-informed guidance sampling. The proposed Pi-fusion can also be generalized in learning other physical dynamics governed by partial differential equations. Data and code are available at https://github.com/SIAT-SIH/fluid.
\end{abstract}


\section{Introduction}

    Understanding the underlying fluid dynamics is essential for exploring complex physical phenomena in a wide range of fields from science to engineering \cite{jhne1994ImagingOS, bateman2001On,sahli2020Cardiac, kissas20201Machine, arzani2021UncoveringNB}. Especially in the medical fields, studying the fluid dynamics of blood flow (hemodynamics) in cardiovascular system is necessary for developing insights into mechanisms of physiology and vascular diseases in microcirculation, which contribute to the prediction and treatment of cardiovascular disease \cite{rassi2019PINN}. Theoretically, the fluid dynamics are governed by partial differential equations (PDEs) such as the well-known Navier-Stokes (N-S) equations. Solving such equations (i.e., learning the solution that maps input variable to the corresponding fluid dynamics characteristics including velocity and pressure field) across different domain geometries and/or input parameters is crucial to simulate fluid dynamics. Even though such equations can be numerically solved by using traditional discretization methods including finite volume \cite{robert2000FVM}, finite element \cite{oc1977FEM} or other approximate methods, they suffer from formidable costs in real-world scenarios (thousands of CPU hours required), as large-scale meshes involved. 
    Recently, several machine learning  approaches, especially deep learning, have been revolutionized our perspective of modeling physical phenomenon and learning fluid dynamics \cite{tracey2015AML,duraisamy2015NewAI, kutz2017DL},
    and they showed promising achievements. 
    Nevertheless, a critical bottleneck of deep learning approaches lies in its strong dependence on the amount and coverage of the training data, which is very difficult, time-consuming and computational expensive to obtain. Hence, the research community has developed a novel paradigm, which incorporates physical mechanisms into machine learning approaches, known as physics-informed deep learning\cite{rassi2019PINN, sirignano2018DGM,lu2021DeepONet,long2018PDENET1}. While physics-informed deep learning have shown early promise in modeling physical phenomenon, simulating real-world fluid dynamics, especially blood flow in the cardiovascular system, are significantly challenging. First, the scale of the system is extremely large (about millions of particles) and the blood flow is complex (e.g., unstable quasiperiodic pattern related to heartbeat cycle), how to ensure the accuracy and generalizability (i.e., the stability in predicting temporal evolution of flow dynamics) of the model remains a critical question. Ideally, one would wish to be able to learn the distribution of the fluid fields. Second, as the PDEs cannot be uncovered in Deep Neural Networks-based model, it remains a “black box” and lacks sufficient interpretability, and thus it often causes controversy when applied in medical problems. The third challenge related to the fact that, the performance of those methods has seldom been verified in real-world scenario, leading to uncertain practical value. However, a maximum likelihood estimation of the loss function via gradient descent methods is usually executed for training in existing physics-informed deep neural network, resulting in any trained neural network to fit a particular type as the training datasets. An efficient approach that is able to learn the distribution of the fluid fields, is thus vital to improve the generalizability and interpretability in predicting fluid dynamics in real-world scenarios.
    
    Diffusion models have recently emerged as a powerful state-of-art series of models. Based on rigorous mathematical interpretability from probability theory perspective, diffusion models have strong capability in learning data distribution and thus achieved record-breaking performance in various applications, including computer vision \cite{dhariwal2021diffusion, kawar2022enhancing, ho2022classifier}, temporal data modeling \cite{chang2023tdstf,li2022generative} and medical imaging \cite{rahman2023ambiguous,bieder2023diffusion}. Diffusion models use Bayesian inference to obtain probabilistic outputs, relying on interpreting each trainable parameter of the network to be a random variable that may be sampled from, which leads to an output that can be characterized with a probability density function.

    Inspired by the fundamental principles of diffusion models, we leverage these advances to propose a physics-informed diffusion model (Pi-fusion), for learning fluid dynamics with incompressible Newtonian flows. By drawing the strengths of diffusion model for learning the data distribution, 
    our approach is with strong generalizability. In particular, our contributions are as follow:
 
    \begin{itemize}
        \item[$\bullet$] We design Pi-fusion, a physics-informed diffusion model, to improve the accuracy of predicting fluid dynamics by learning the data distribution across different temporal instants. The proposed model also possess interpretability by involving physical mechanism in the guidance sampling. 
        \item[$\bullet$] We propose a training strategy based on reciprocal learning, to explore the quasiperiodic pattern of fluid motion and thus further enhance the generalizability in predicting the temporal evolution of velocity and pressure field.
        \item[$\bullet$] Our approach is verified on both synthetic and real-world problems, to demonstrate its performance to real, complex domain constrained data.
    \end{itemize}

\section{Background}
\subsection{Incompressible fluid flow}\label{2.1}
    
\begin{figure}[t]
		\setlength{\belowcaptionskip}{-0.2cm}
		\centering
		\includegraphics[width=0.8\linewidth]{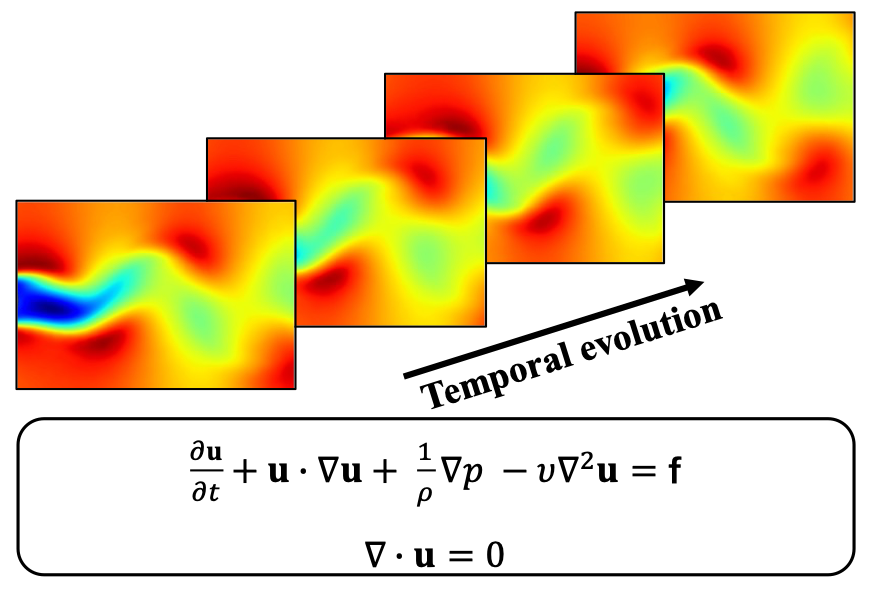}
            \vspace{-5pt}
		\caption{Visualization of time evolution of the velocity in the case of 2D Compressible NS equations.}
		\label{characteristics}
    \end{figure}

     Our work mainly considers incompressible flow governed by the N-S equations, which can be expressed as:
    \begin{equation}
        \frac{\partial\mathbf{u}}{\partial t}+\mathbf{u}\cdot \nabla \mathbf{u} + \frac{1}{\rho}\nabla p-\nu\nabla^2\mathbf{u}= \mathbf{f} 
        \quad\mathrm{momentum \ conservation},
    \label{eq:ns1}
    \end{equation}
    \begin{equation}
        \nabla \cdot \mathbf{u} = 0 \quad\quad\quad\quad\quad\quad\quad\mathrm{incompressibility}
    \label{eq:ns2}
    \end{equation}
    where $\mathbf{u}\in \mathbb{R}^n$ is the velocity (vector), $p$ denotes the pressure (scalar), $t$ is time, $\rho$ represents the density, $\nu$ is the kinematic viscosity (assumed to be constant), 
    and $\textbf{f}$ is the body force. In this work, we set $\textbf{f} = 0$ since external forces such as gravity can be neglected and geometry refers to the coordinates of points. As shown in Fig.~\ref{characteristics}, the flow characteristics is extremely complex because fluid systems exhibit temporal-varying behaviors. Existing deep learning methods were merely applied to learn the flow fields of the same time instants as the training data, which have difficulty in generalizing to arbitrary temporal instants.
    How to predict the temporal evolution of flow field is a critical issue, yet has not addressed in existing deep learning methods.

\subsection{Learning-based methods for modeling physical dynamics}
    Although several learning-based surrogate models to tackle physics problems have been proposed, from deep learning methods \cite{sirignano2018DGM, long2018PDENET1} to Physics Informed Neural Networks (PINN) \cite{rassi2019PINN, lim2022PIFD, fang2022HPIN} and neural operators \cite{kovachki2021NO, Li2020FNO, li2020Multipole, lu2021DeepONet}, reliably inferring fluid velocity and pressure fields remains a challenging task. Most of the existing convolutional neural network-based studies for solving problems represented by N-S equations \cite{nils2020Teaching, mohan2023Embedding, wandel2020LearningIF, wang2020Towards} only can handle regular geometries with uniform grids, rendering them unable to tackle non-uniform observations commonly encountered in realistic application. Other models such as \cite{Li2020FNO} lie on a regular grid to perform a Fast Fourier Transform of the input data. Such neural operators are not designed to directly operate on unstructured data, resulting in inaccurate predictions of the physical fields at the surface of geometries. 
    The auto-regressive inference methods in the literature like \cite{zhu2019Pcdl, geneva2020Modeling} suffer from error accumulation, which is particularly severe for the long-term formulation. 
    While works like \cite{raissi2020Hidden, jin2021NSFnets, kashefi2022PointNet} focused on the simulation of the incompressible N-S equations in determinant manner, they face challenges in learning the underlying data distribution, posing difficulties in generalization to diverse temporal and spatial geometries. Moreover, these methods has seldom been verified in real-world scenario, leading to uncertain practical value in dealing with a broader range of scenarios and dimensional complexity.

\subsection{Diffusion models}
    Diffusion models are a type of generative models that have recently gained attention in many fields. They are inspired by non-equilibrium thermodynamics and learn to reverse the forward process of sequentially corrupting data samples with additive random noise, until reconstructing the desired data that matches the source data distribution from noise. There are two main types of diffusion models, diffusion-based \cite{ho2020denoising} and score-matching based \cite{hyvarinen2005estimation, vincent2011connection}. Following them, denoising diffusion probabilistic models \cite{ho2020denoising, nichol2021improved} and noise-conditioned score networks \cite{song2019generative, song2020score, song2020improved,meng2021sdedit, chung2022come} are proposed to synthesize high-quality images, respectively. Diffusion-based models show great potential in various tasks such as image synthesis \cite{dhariwal2021diffusion, kawar2022enhancing, ho2022classifier}, inpainting \cite{lugmayr2022repaint}, super-resolution \cite{lee2022progressive, saharia2022image}, deblurring \cite{whang2022deblurring}, and image-to-image translation \cite{wang2022pretraining, saharia2022palette, choi2021ilvr}. There are many methods\cite{chung2022diffusion,chung2022improving,chung2022parallel,wang2022zero} to solve the inverse problems, but they are designed to some special degradation model.  In summary, in view of the powerful distribution fitting ability and interpretability of diffusion models, it is promising to exploit it into physical-informed deep learning to model the  spatio-temporal fluid dynamics. Even though a physics-informed diffusion model was proposed to reconstruct high-fidelity flow data\cite{dule2023pidm}, there is no framework designed for learning fluid dynamics based on diffusion models. The basic principles of diffusion models are presented as follows.
\subsubsection{Denoising diffusion probabilistic models}\label{3.4}
	
	Denoising Diffusion Probabilistic Models (DDPMs) \cite{ho2020denoising, nichol2021improved} destroy training data by adding Gaussian noise, and then recover the data by reverse process. The forward process follows the Markov chain that transforms a data sample $\x_0\sim q(\x_0)$ into a sequence of noisy samples $\x_t$ in $T$ steps with a variance schedule $\beta_1,\ldots,\beta_T$:
	\begin{equation}
		q(\x_{t}\vert\x_{t-1}) = \mathcal{N}(\x_{t};\sqrt{1-\beta_t}\x_{t-1},\beta_t \I),
		\label{eq:forward1}
	\end{equation}
	
	The above process can be reversed by $p_{\theta}(\x_{0:T})$ with learnable parameters $\theta$, starting from standard Gaussian distribution $p(\x_T)=\mathcal{N}(\x_T;\mathbf{0},\I)$:
	\begin{equation}
		p_{\theta}(\x_{t-1}\vert\x_t) = \mathcal{N}(\x_{t-1};\bm{\mu}_{\theta}(\x_t,t),\mathbf{\Sigma}_{\theta}(\x_t,t)).
		\label{eq:reverse}
	\end{equation}
	$p_{\theta}(\x_{0:T})$ is a neural network that predicts $\bm{\mu}_{\theta}(\x_t,t)$ and $\mathbf{\Sigma}_{\theta}(\x_t,t)$. As reported by \cite{ho2020denoising}, this optimization can be converted to train a network $\bm{\epsilon}_{\theta}(\x_t,t)$ to predict the noise vector by the reparameterization of the reverse process:
	\begin{equation}
		\bm{\mu}_{\theta}(\x_t,t) = \frac{1}{\sqrt{\alpha_t}} \left(\x_t - \frac{\beta_t}{\sqrt{1-\bar{\alpha}_t}}\bm{\epsilon}_{\theta}(\x_t,t)\right),
		\label{eq:reparameterization}
	\end{equation}
	where $\alpha_t=1-\beta_t$, $\bar{\alpha}_t=\prod_{i=1}^t\alpha_i$.
	As a result, the training objective is transformed into a re-weighted simplified form given as:
	\begin{equation}
		L_{simple} =  \mathbb{E}_{\x_0,t,\bm{\epsilon}_t\sim\N(\mathbf{0},\I)}\Big[\vert\vert\bm{\epsilon}_t -  \bm{\epsilon}_{\theta}(\x_t,t)\vert\vert^2 \Big].
		\label{eq:training_obj1}
	\end{equation}
	Then the sampling $p_{\theta}(\x_{t-1}\vert\x_t)$ starts from $\x_T\sim\N(\mathbf{0},\I)$ and adds noise $\bm{z}\sim\mathcal{N}(\mathbf{0},\I)$ in each iteration:
	\begin{equation}
		\x_{t-1}=\frac{1}{\sqrt{\alpha_t}}\left(\x_t - \frac{\beta_t}{\sqrt{1-\bar{\alpha}_t}} \bm{\epsilon}_{\theta}(\x_t,t)\right) + \sigma_t\bm{z}.
	\end{equation}
	
	\subsubsection{Deterministic implicit sampling}\label{3.5}
	
	Denoising Diffusion Implicit Models (DDIMs) \cite{song2020denoising} have the same training objective as Eq. \ref{eq:training_obj1} by defining a non-Markovian diffusion process:
	\begin{equation}
		q_{\sigma}(\x_{t}\vert\x_0)=\N(\sqrt{\bar{\alpha}_{T}}\x_{0};(1-\sqrt{\bar{\alpha}_{T}})\I), 
	\end{equation}
	\begin{equation}
		q_{\sigma}(\x_{t-1}\vert\x_t,\x_0)=\N(\x_{t-1};\bm{\Tilde{\mu}}_t(\x_t,\x_0),{\sigma}_t^2\I).
		\label{ddim sampling}
	\end{equation}
	By setting $\sigma_t^2=\frac{1-\bar{\alpha}_{t-1}}{1-\bar{\alpha}_t}\beta_t$, the forward process becomes Markovian and remains the same as DDPMs.
	
	A deterministic implicit sampling is implemented by setting $\sigma_t^2=0$, and thus the reverse process based on Eq. \ref{ddim sampling} is rewritten by: 
	\begin{equation}
		\begin{split}
			\x_{t-1} &=  \sqrt{\bar{\alpha}_{t-1}}\left(\frac{\x_t-\sqrt{1-\bar{\alpha}_t}\cdot\bm{\epsilon}_{\theta}(\x_t,t)}{\sqrt{\bar{\alpha}_t}}\right) \\ & + \sqrt{1-\bar{\alpha}_{t-1}}\cdot\bm{\epsilon}_{\theta}(\x_t,t),
		\end{split}
		\label{eq:ddim}
	\end{equation}
	which allows a faster sampling by using only the subsequence $\x_T, \x_{\tau_S},\x_{\tau_{S-1}},\ldots,\x_{\tau_1}$ with $\tau_i = (i-1)\cdot T / S$, where $S < T$ is the number of sampling steps.

\section{Method}

\begin{figure*}[t]
\begin{minipage}[b]{1\linewidth}
    \centering
    \subfloat[]{\includegraphics[width=0.98\linewidth]{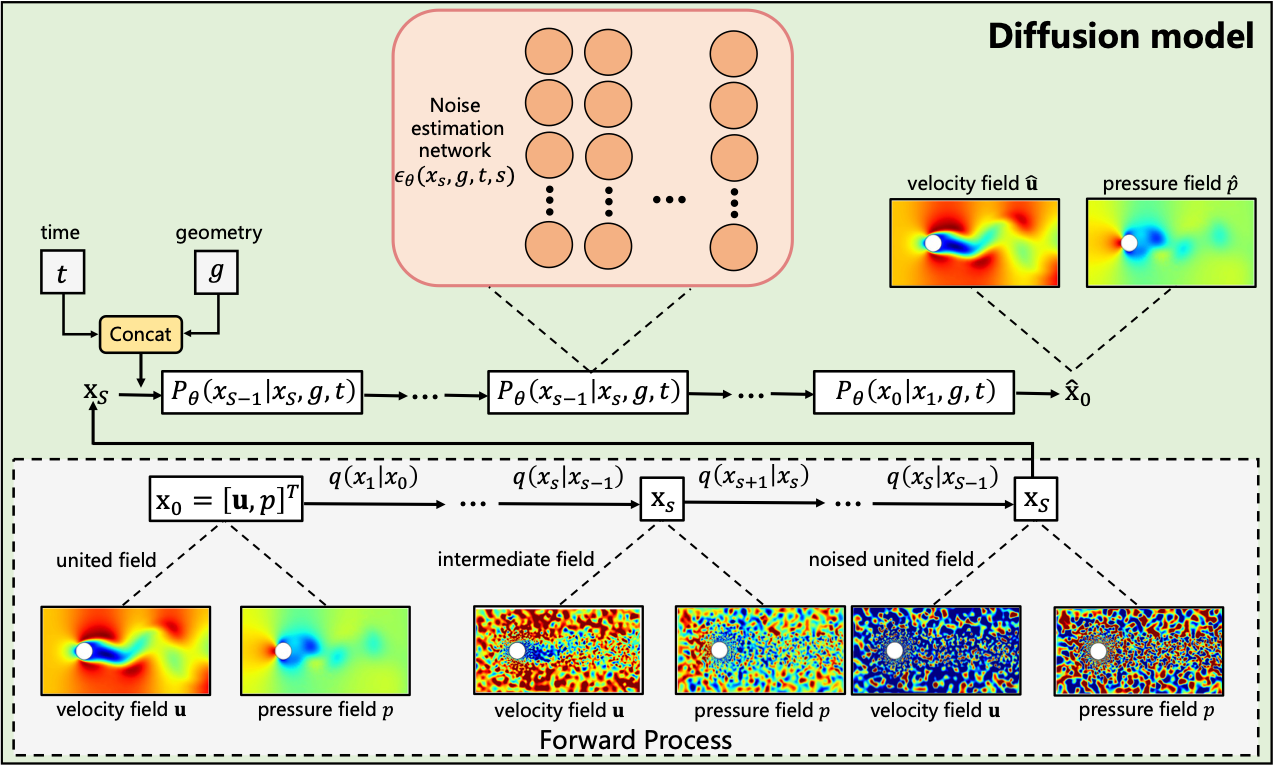}}
    \vspace{-5pt}
\end{minipage}
\begin{minipage}[b]{1\linewidth}
    \centering
    \subfloat[]{\label{figtr}\includegraphics[width=0.485\linewidth]{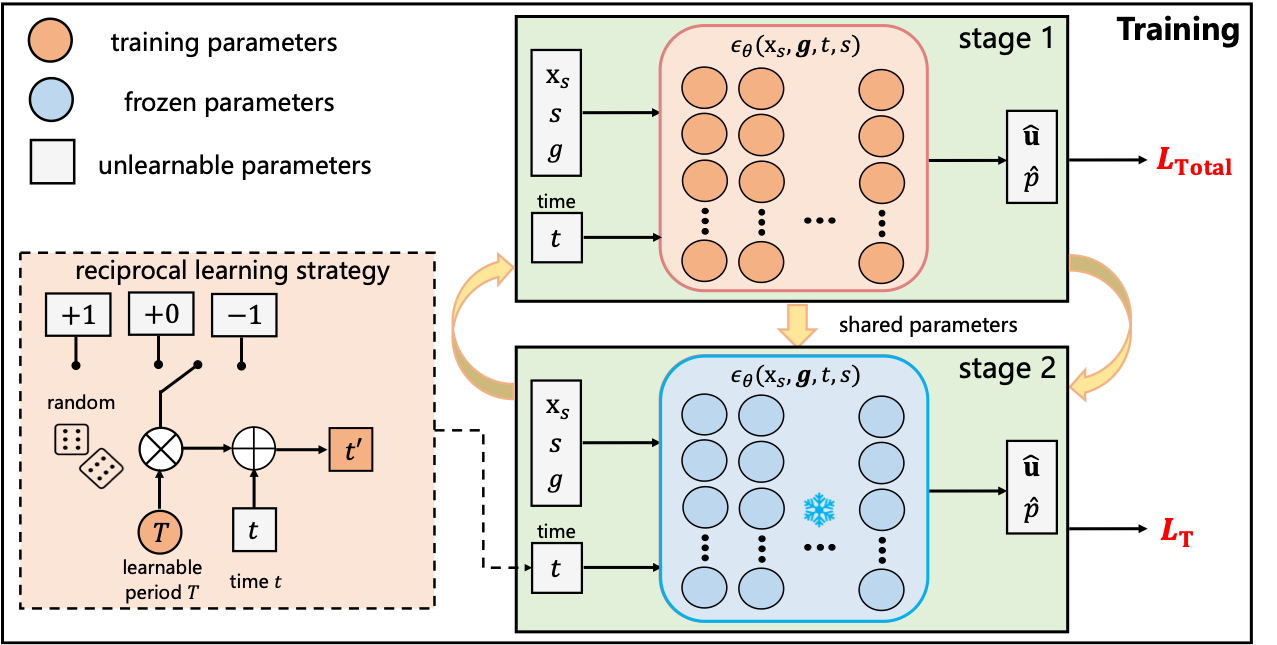}}
    \subfloat[]{\label{figsa}\includegraphics[width=0.495\linewidth]{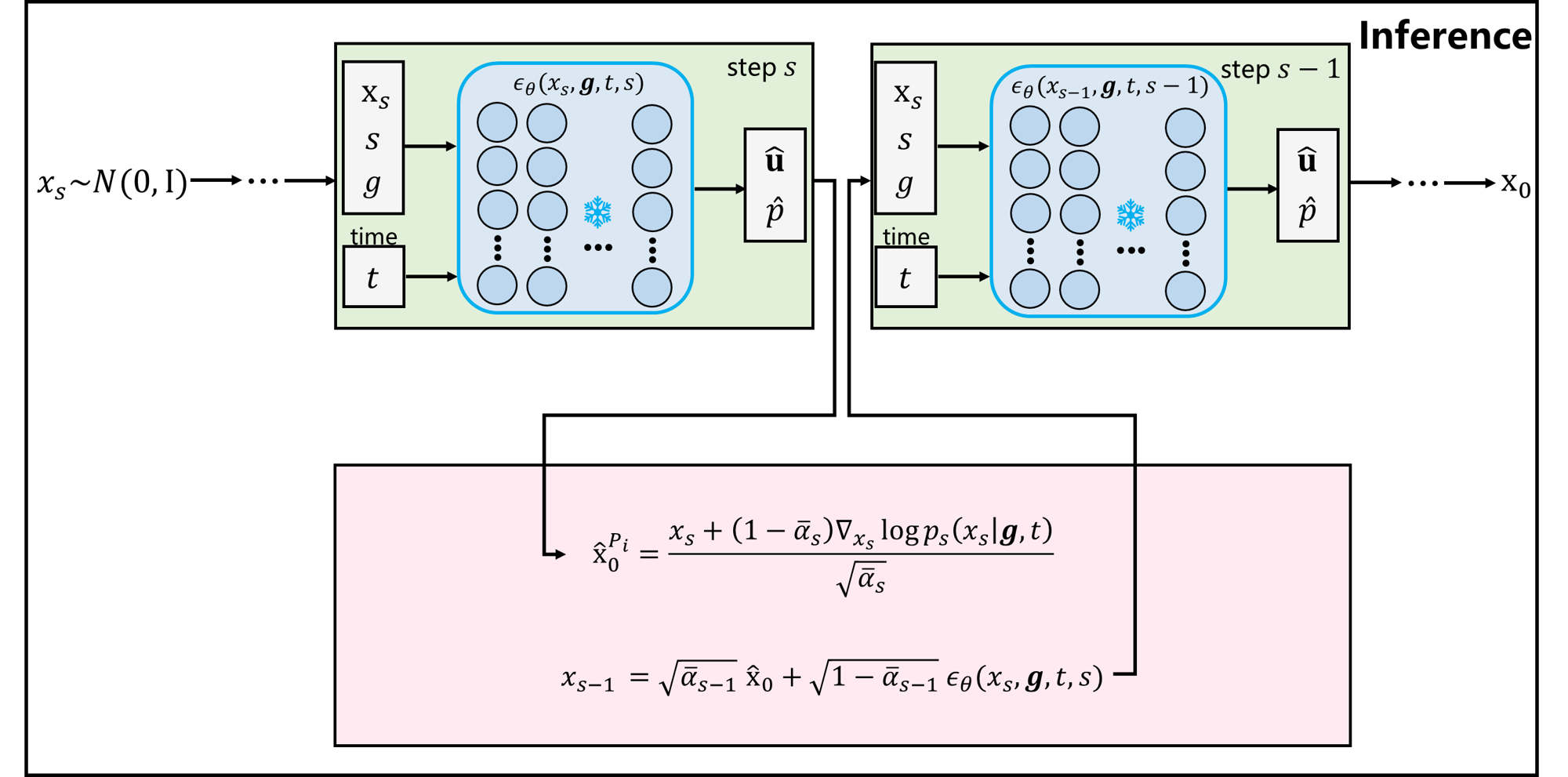}}
\end{minipage}
\vspace{-3pt}
\caption{(a) The architecture of diffusion model in Pi-fusion. (b) The reciprocal learning strategy in the training of Pi-fusion. (c) The physics-informed guidance sampling in the inference of Pi-fusion. }
\label{fig:all}
\vspace{-9pt}
\end{figure*}




\subsection{Problem formulation}\label{4.1}
We first give the detailed notation of the model formulation as follows, with 2D incompressible flow as the example. Given 2D velocity field $\mathbf{u}=[u,v]^\top$, 2D geometry $\mathbf{g} = [x,y]^\top$ and the pressure field $p$, we can rewrite Eq. \ref{eq:ns1} and \ref{eq:ns2} as
\begin{equation}\label{pde1}
\frac{\partial u}{\partial t}+u \frac{\partial u}{\partial x}+v \frac{\partial u}{\partial y}=-\frac{1}{\rho} \frac{\partial p}{\partial x}+\nu\left(\frac{\partial^{2} u}{\partial x^{2}}+\frac{\partial^{2} u}{\partial y^{2}}\right),
\end{equation}
\begin{equation}\label{pde2}
\frac{\partial v}{\partial t}+u \frac{\partial v}{\partial x}+v \frac{\partial v}{\partial y}=-\frac{1}{\rho} \frac{\partial p}{\partial y}+\nu\left(\frac{\partial^{2} v}{\partial x^{2}}+\frac{\partial^{2} v}{\partial y^{2}}\right),
\end{equation}
\begin{equation}\label{pde3}
\frac{\partial u}{\partial x}+\frac{\partial v}{\partial y}=0.
\end{equation}
For the above problem, the specific task is to predict the 2D velocity field $\mathbf{u}$ and pressure field $p$ given the certain 2D geometries $\mathbf{g}$ and time $t$ as follows:
\begin{equation}
\mathbf{u}, p = \argmax_{\mathbf{u},p} q(\mathbf{u},p\ |\ \mathbf{g}, t),  
\end{equation}
where $q(\mathbf{u},p\ |\ \mathbf{g}, t)$ is the posterior distribution of $\mathbf{u}$ and $p$ conditioned on $\mathbf{x}$ and $t$. Diffusion model is employed to better fit the learning of this posterior distribution. According to the principles of diffusion model presented in \ref{3.4}, the forward process in DDPMs aims to learn the distribution of $\x_0$, i.e., $\mathbf{u}$ and $p$ (the velocity and pressure fields).  

 We denote $\x_0 = [\mathbf{u},p]^\top$ as the united field of velocity and pressure. We use $s$ and $S$ to express the step and the number of steps in the diffusion model. Specifically, the forward process of Pi-fusion can be described as follows:
    \vspace{-3pt}
    \begin{equation}
		q(\x_{s}|\x_{s-1}) = \mathcal{N}(\x_{s}; \bm{\mu}_s(\x_{s-1}),(1-\alpha_s)\I),
		\label{eq:forward2}
	\end{equation}
    \begin{equation}\label{eqforward}
    q(\x_{s}\vert\x_0)
    =\N(\x_s;[\sqrt{\bar{\alpha}_s}\mathbf{u},\sqrt{\bar{\alpha}_s}p]^\top,(1-\bar{\alpha}_s)\I).
	\end{equation}
 
With $\epsilon_s \sim \N(0,\I)$, we can rewrite the above forward process as follows:
\begin{equation}\label{eqforwards}
    \x_{s}=\sqrt{\bar{\alpha}_s}\x_0+\sqrt{1-\bar{\alpha}_s}\epsilon_s=[\sqrt{\bar{\alpha}_s}\mathbf{u},\sqrt{\bar{\alpha}_s}p]^\top+\sqrt{1-\bar{\alpha}_s}\epsilon_s.
	\end{equation}
Following DDIMs presented in \ref{3.5} \cite{song2020denoising}, we can derive the reverse process from $\x_S$ to $\x_0$ by:
    \begin{equation}
    \begin{aligned}
		q(\x_{s-1}\vert\x_s,\x_0)=\N(\x_{s-1};\bm{\Tilde{\mu}}_s(\x_s,\x_0),\Tilde{\sigma}_s^2\I),
  \end{aligned}
		\label{DDIM sampling}
	\end{equation}
 \begin{equation}
        \begin{split}
		\bm{\Tilde{\mu}}_s&=\sqrt{1-\bar{\alpha}_{s-1}}\frac{\x_s-\sqrt{\bar{\alpha}_s}\x_0}{\sqrt{1-\bar{\alpha}_s}}+\sqrt{\bar{\alpha}_{s-1}}\x_0,
		\end{split}
		\label{DDIM mean}
	\end{equation}
    where $\x_0$ can be predicted by a noise estimation network $\bm{\epsilon}_{\theta}(\x_s,\mathbf{g},t,s)$:
    \begin{equation}
        \begin{split}
		\bm{\hat{\x}}_0&=\frac{\x_s-\sqrt{1-\bar{\alpha}_s}\bm{\epsilon}_{\theta}(\x_s,\mathbf{g},t,s)}{\sqrt{\bar{\alpha}_s}}.
		\end{split}
		\label{DDIMx0}
	\end{equation}
The architecture of the proposed Pi-fusion is presented in Fig.~\ref{fig:all} (a), a multilayer perceptron (MLP) architecture is used as the noise estimation network $\bm{\epsilon}_{\theta}(\x_s,\mathbf{g},t,s)$ with the adjustment of the input and the output size to adapt to learn the fluid dynamics. 

In each step $s$, the given geometries $\mathbf{g}$ and the time $t$ serve as the conditions in Pi-fusion to model the mixed distribution of the united field $[\mathbf{u},p]^\top$. 
    The base loss function of diffusion models $\mathcal{L}_{simple}$ is utilized as the main supervision for $\bm{\epsilon}_{\theta}(\x_s,\mathbf{g},t,s)$:
    \begin{equation}
		\mathcal{L}_{simple} =  \mathbb{E}_{\x_0,s,\bm{\epsilon}_s\sim\N(\mathbf{0},\I)}\Big[\vert\vert\bm{\epsilon}_s -  \bm{\epsilon}_{\theta}(\x_s,\mathbf{g},t,s)\vert\vert^2 \Big].
		\label{eq:training_obj}
	\end{equation}
    
    After obtaining $\hat{\x}_0$ in Eq. \ref{DDIMx0}, we have $\hat{\x}_0=[\hat{\mathbf{u}},\hat{p}]^\top=[\hat{u},\hat{v},\hat{p}]^\top$. Then, according to the Eqs.~\ref{pde1}--\ref{pde3}, the following PDEs loss is also included for the training of the $\bm{\epsilon}_{\theta}(\x_s,\mathbf{g},t,s)$: 
    \begin{equation}\label{pde4}
    L_u = \frac{\partial \hat{u}}{\partial t}+\hat{u} \frac{\partial \hat{u}}{\partial x}+\hat{v} \frac{\partial \hat{u}}{\partial y}+\frac{1}{\rho} \frac{\partial \hat{p}}{\partial x}-\nu\left(\frac{\partial^{2} \hat{u}}{\partial x^{2}}+\frac{\partial^{2} \hat{u}}{\partial y^{2}}\right),
    \end{equation}
    \begin{equation}\label{pde5}
    L_v =\frac{\partial \hat{v}}{\partial t}+\hat{u} \frac{\partial \hat{v}}{\partial x}+\hat{v} \frac{\partial \hat{v}}{\partial y}+\frac{1}{\rho} \frac{\partial \hat{p}}{\partial y}-\nu\left(\frac{\partial^{2} \hat{v}}{\partial x^{2}}+\frac{\partial^{2} \hat{v}}{\partial y^{2}}\right),
    \end{equation}
    \begin{equation}\label{pde6}
    L_{\mathbf{g}} = \frac{\partial \hat{u}}{\partial x}+\frac{\partial \hat{v}}{\partial y},
\end{equation}
    \begin{equation}\label{pdeloss}
    \begin{aligned}
        \mathcal{L}_{\mathrm{PDEs}}(\hat{\x}_0,\mathbf{g},t) 
        & = ||L_u-0||_2+||L_v-0||_2+||L_{\mathbf{g}}-0||_2,\\
    \end{aligned}
    \end{equation}
    where $\mathbf{g} = [x,y]^\top$. Finally, by including data loss $\mathcal{L}_{\mathrm{data}}  = ||\hat{\mathbf{u}}-\mathbf{u}||_2+||\hat{p}-p||_2$, we obtain:
    \begin{equation}\label{allloss}
    \begin{aligned}
        \mathcal{L}_{\mathrm{Total}} 
        & = \mathcal{L}_{\mathrm{data}} + \lambda_{PDEs}\mathcal{L}_{\mathrm{PDEs}} + \lambda_{simple}\mathcal{L}_{simple}.\\
    \end{aligned}
    \end{equation}
    We set the weights $\lambda_{PDEs}$ and $\lambda_{simple}$ following the process presented in \cite{nair2022ddpm}.
    
\subsection{Training strategy based on reciprocal learning}\label{4.2}
Usually, the incompressible fluid flow has periodic behavior, especially for blood flow in the vasculature, the period is unstable and directly related to the heartbeat cycle. Knowing the periodic behavior of the flow field is of great importance for learning the fluid dynamics. To further enhance the generalizability of the proposed Pi-fusion, we proposed a training strategy based on reciprocal learning to explore the periodic behavior in fluid motion. For fluid flow with a period $T$, we can get $q(\mathbf{u},p\ |\ \mathbf{g}, t+T) = q(\mathbf{u},p\ |\ \mathbf{g}, t)$ and $q(\mathbf{u},p\ |\ \mathbf{g}, t-T) = q(\mathbf{u},p\ |\ \mathbf{g}, t)$ with the propose Pi-fusion. Therefore, we inspiringly propose a strategy for period learning, as presented in Fig.~\ref{fig:all} (b). 

This strategy is divided into two alternating stages. In the stage 1, $\bm{\epsilon}_{\theta}(\x_s,\mathbf{g},t,s)$ is trained to predict the united field. After $\bm{\epsilon}_{\theta}(\x_s,\mathbf{g},t,s)$ is adequately trained (loss is less than a fixed threshold), the parameters are transferred and then train the period parameter $T$ to represent the period. When the loss to train period $T$ is stable, we then repeat the stage 1 and so on until both stages converge. The loss function $\mathcal{L}_{T}$ is utilized as the supervision for learning $T$, and we set the time $t^{\prime}$ input into $\bm{\epsilon}_{\theta}$ as $t^{\prime}=t+\lambda T$, where $\lambda$ is obtained by random sampling from $\{-1, 0, 1\}$.
    \begin{equation}
		\mathcal{L}_{T} =  \mathbb{E}_{\x_0,s,\bm{\epsilon}_s\sim\N(\mathbf{0},\I),\lambda\in \{-1,0,1\}}\Big[\vert\vert\bm{\epsilon}_s -  \bm{\epsilon}_{\theta}(\x_s,\mathbf{g},t^{\prime},s)\vert\vert^2 \Big].
		\label{eq:training_obj2}
	\end{equation}
According to the training strategy presented above, if $\bm{\epsilon}_{\theta}(\x_s,\mathbf{g},t,s)$ is trained adequately in stage 1, it can help stage 2 learn a more accurate period $T$. Similarly, if the $T$ is trained appropriately in stage 2, it can in turn help improve the generalization of $\bm{\epsilon}_{\theta}(\x_s,\mathbf{g},t,s)$ in stage 1. Through this reciprocal learning, the proposed Pi-fusion can learn the fluid dynamics more accurately.

\begin{algorithm}[t] 
	\caption{\textbf{: Pi-fusion's training strategy}} 
	\label{algTrain} 
	\begin{algorithmic}[1] 
		  \While {not converged}
		\State 2D geometry $\mathbf{g}$, time $t$, velocity field $\mathbf{u}$, and the pressure field $p$ from the dataset. Let $\x_0 = [\mathbf{u},p]^\top$;
            \For {training stage 1}   
		\State Step $s \sim \mathbf{Uniform}({1,..S})$; $\bm{\epsilon}\sim \mathcal{N}(\bm{0},\I)$;
		\State $\x_{s}=\sqrt{\bar{\alpha}_s}\x_0+\sqrt{1-\bar{\alpha}_s}\epsilon$ as Eq. \ref{eqforwards};\quad $t^{\prime}=t$;
        \If {stage 2 converged}
        \State With the learned period $T$, get $t^{\prime}=t+\lambda T$ randomly, $\lambda \in \{-1, 0, 1\}$;
        \EndIf
        \State Obtain $\bm{\hat{\x}}_0=(\x_s-\sqrt{1-\bar{\alpha}_s}\bm{\epsilon}_{\theta}(\x_s,\mathbf{g},t,s))/\sqrt{\bar{\alpha}_s}$ as Eq.~\ref{DDIMx0};
            \State Take gradient descent $\nabla_{\bm{\epsilon}_\theta}\mathcal{L}_{\mathrm{Total}}$ for the learning of $\bm{\epsilon}\theta$;
            \EndFor
            \For {training stage 2}   
		\State Step $s \sim \mathbf{Uniform}({1,..S})$; $\bm{\epsilon}\sim \mathcal{N}(\bm{0},\I)$;
		\State $\x_{s}=\sqrt{\bar{\alpha}_s}\x_0+\sqrt{1-\bar{\alpha}_s}\epsilon$ as Eq. \ref{eqforwards};
        \State With the learnable period $T$, get $t^{\prime}=t+\lambda T$ randomly, $\lambda \in \{-1, 0, 1\}$.
        \State Obtain $\bm{\hat{\x}}_0=(\x_s-\sqrt{1-\bar{\alpha}_s}\bm{\epsilon}_{\theta}(\x_s,\mathbf{g},t^{\prime},s))/\sqrt{\bar{\alpha}_s}$ as Eq.~\ref{DDIMx0};
            \State Take gradient descent $\nabla_{T}\mathcal{L}_{T}$ for the learning of the learnable period $T$;
            \EndFor
		\EndWhile
	\end{algorithmic} 
\end{algorithm}

  \begin{algorithm}[t] 
	\caption{\textbf{: Pi-fusion's physics-informed guidance sampling}} 
	\label{algPGPS} 
	\begin{algorithmic}[1] 
            \Require 2D geometry $\mathbf{g}$, time $t$
            \State $\x_S \sim \mathcal{N}(0,\I)$.
		  \For {$s=S,..,1$}   
		\State Compute $\bm{\epsilon}_{\theta}(\x_s,\mathbf{g},t,s)$;
            \State $\hat{\x}^{\mathrm{Pi}}_0=\frac{\x_s+(1-\bar{\alpha}_s)\nabla_{\boldsymbol{\x}_{s}} \log p_{s}\left(\boldsymbol{\x}_{s}|\mathbf{g},t\right) }{\sqrt{\bar{\alpha}_s}})$;
            \State Obtain $\x_{s-1}$ by Eq.~\ref{eqxnext} with $\hat{\x}_0$ replaced by $\hat{\x}^{\mathrm{Pi}}_0$;
		\EndFor
            \Ensure $\hat{\x}_0 = [\mathbf{\hat{u}},\hat{p}]^\top$.
	\end{algorithmic} 
\end{algorithm} 

\subsection{Physics-informed guidance sampling for model inference}\label{4.3}

   According to Eqs.~\ref{DDIM sampling}--\ref{DDIMx0}, the vanilla sampling can be reparameterized with $\bm{\hat{\x}}_0$ and Eq.~\ref{DDIM mean}:
 \begin{equation}\label{eqxnext}
 \begin{split}
 &\x_{s-1}=\sqrt{\bar{\alpha}_{s-1}}\hat{\x}_0+\sqrt{1-\bar{\alpha}_{s-1}}\cdot\bm{\epsilon}_{\theta}(\x_s,\mathbf{g},t,s).
 \end{split}
 \end{equation}
 
By considering the PDEs in Eqs.~\ref{pde1}--\ref{pde3}, we propose a physics-informed guidance sampling for model inference. With the Bayes' rule, we have: 
 \begin{equation}
 \begin{aligned}
p(\x_s|\mathbf{g},t)=p(\mathbf{g},t|\x_s)p(\x_s)/p(\mathbf{g},t)
 \end{aligned}
    \end{equation}
    
    \begin{equation}\label{eqgrad}
    \begin{aligned}
        \nabla_{\boldsymbol{x}_{s}}\log p_{s}(\x_{s}|\mathbf{g},t)=\underbrace{\nabla_{\boldsymbol{x}_{s}} \log p_{s}\left(\mathbf{g},t|\boldsymbol{x}_{s}\right)}_{\mathrm{PDEs}}+\underbrace{\nabla_{\x_{s}} \log p_{s}\left(\boldsymbol{x}_{s}\right)}_{\mathrm{United\ field}},
    \end{aligned}
    \end{equation}
    where $\nabla_{\boldsymbol{x}_{s}} \log p_{s}(\x_{s}|\mathbf{g},t)$ is the posterior score function denoting the gradient direction to predict a $\x_s$, with which we can sample a $\x_s$ that is more consistent with the distribution $p(\x_s|\mathbf{g},t)$. Next, we derive the closed forms of the PDEs term and the united field term.

    From Eq.~\ref{pdeloss}, we have that the ground truth united field and geometry should conform to $\mathcal{L}_{\mathrm{PDEs}}(\x_0,\mathbf{g},t)=0$. So we assume that there is a complex implicit function $\phi([\mathbf{g},t]^\top) = [\mathbf{u}, p]^\top = \x_0$ and its inverse function $[\mathbf{g},t]^\top =\phi^{-1}(\x_0)$, which makes sure that $\mathcal{L}_{\mathrm{PDEs}}(\x_0,\mathbf{g},t)=0$.
    So it is reasonable to assume the conditional distribution $p\left(\mathbf{g},t|\boldsymbol{\x}_{0}\right) \simeq \mathcal{N}\left(\mathbf{g},t;\phi^{-1}(\x_0), \sigma^{2}_{\mathrm{PDE}} \I\right)$. 
   
   Based on Eq.~\ref{DDIMx0}, we can have the approximation $p_{s}\left(\mathbf{\x}_{0}|\mathbf{x}_{s}\right) \simeq \mathcal{N}\left(\x_0;\hat{\mathbf{x}}_{0}, d_{s}^{2} \I\right)$. Thus, we obtain $p_{s}\left(\mathbf{g},t|\mathbf{x}_{s}\right) \simeq \mathcal{N}(\mathbf{g},t;\phi^{-1}(\hat{\mathbf{x}}_{0}),\sigma^{2}_{\mathrm{PDE}}\I)$, with which we can have the following: 
   \begin{equation}
    \begin{aligned}
   &p_{s}\left(\mathbf{g},t|\mathbf{x}_{s}\right) \simeq \mathcal{N}(\mathbf{g},t;\phi^{-1}(\hat{\mathbf{x}}_{0}),\sigma^{2}_{\mathrm{PDE}}\I) =\\
   \\
   & = \frac{1}{\sqrt{(2\pi)^{n+1} |\sigma^{2}_{\mathrm{PDE}}\I|}}\exp(-\frac{\mathcal{L}_{\mathrm{PDEs}}(\hat{\x}_0,\mathbf{g},t)^2}{2\sigma^{2}_{\mathrm{PDE}}}).
   \end{aligned}
    \end{equation}
    Then we can derive that
    \begin{equation}
    \begin{aligned}
        \nabla_{\boldsymbol{\x}_{s}} \log p_{s}\left(\boldsymbol{\x}_{s}|\mathbf{g},t\right)&= \nabla_{\boldsymbol{x}_{s}} \log p_{s}\left(\mathbf{g},t|\boldsymbol{x}_{s}\right)+ s_{\theta}\\
        &= -\frac{\mathcal{L}_{\mathrm{PDEs}}(\hat{\x}_0,\mathbf{g},t)}{\sigma^{2}_{\mathrm{PDE}}}\frac{\partial \mathcal{L}_{\mathrm{PDEs}}(\hat{\x}_0,\mathbf{g},t)}{\partial \x_s}+ s_{\theta},
    \end{aligned}
    \end{equation}
	where $s_{\theta}=\nabla_{\x_{s}}\log p_{s}(\boldsymbol{\x}_{s})=-\bm{\epsilon}_{\theta}/\sqrt{1-\bar{\alpha}_s}$ is the score function derived from the score matching stochastic differential equation\cite{song2020score}. To predict $\x_0$ from $p_{s}(\boldsymbol{\x}_{s}|\mathbf{g},t)$, we need to know the relationship between them. Similar to the derivation in \cite{chung2022diffusion}, we can derive the following result:
    \begin{equation}\label{eqnewx0}
        \begin{aligned}
        \hat{\x}^{\mathrm{Pi}}_0=&\frac{\x_s+(1-\bar{\alpha}_s)\nabla_{\boldsymbol{\x}_{s}} \log p_{s}\left(\boldsymbol{\x}_{s}|\mathbf{g},t\right) }{\sqrt{\bar{\alpha}_s}},\\
    \end{aligned}
    \end{equation}
     The training algorithm of Pi-fusion is given in Algorithm \ref{algTrain}, and the sampling procedure is shown in Algorithm \ref{algPGPS}.
    
    For the input of step number in diffusion model, the corresponding embedding is obtained from the embedding layer according to the integer value of the step. The noised united filed $\x_s$, time $t$ and geometry $\mathbf{g}$ of the input are then passed through different layers of Residual MLPs to extract the respective features, and the different physically meaningful information is mapped onto the same high dimensional feature space. The features from different sources are multiplied by their learnable weight coefficients and then passed through the next 10 layers of Residual MLPs to produce a prediction of the united field. The learning rate is initialized to be 0.001 and decreases for every 200 iterations. We set $\lambda_{PDEs} = 0.1$, $\lambda_{simple}=0.5$ and the number of step $S = 1000$.


 \begin{figure}[t]
		\setlength{\belowcaptionskip}{-0.2cm}
		\centering
		\includegraphics[width=\linewidth]{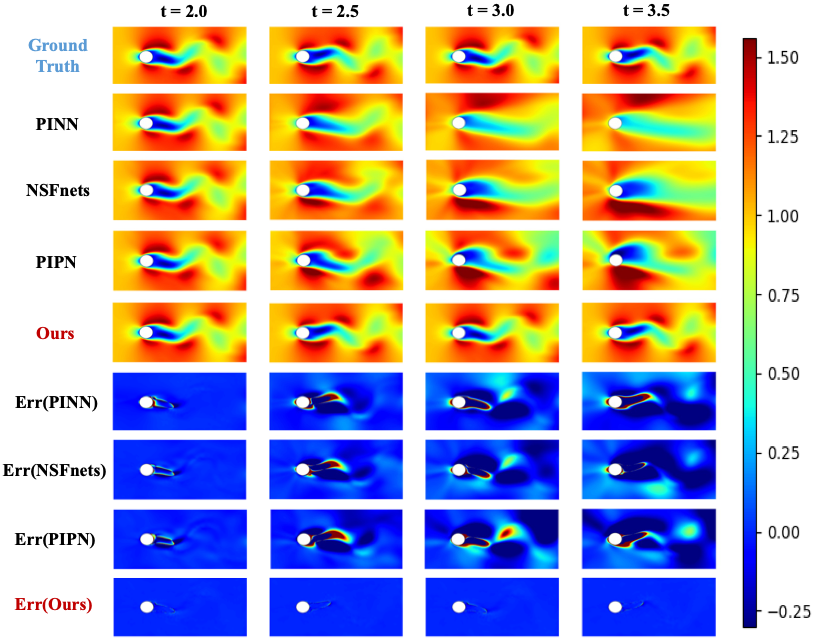}
            \vspace{-15pt}
		\caption{Comparison of the accuracy between Pi-fusion and other methods for the synthetic data in terms of x-direction velocity. Four representative time snapshots are chosen as the example ($t = 2.0s, 2.5s, 3.0s, 3.5s$). Err refers to the difference in the entire domain between ground truth (using direct numerical simulation) and prediction by the approach.}
		\label{result_u}
		\vspace{-0.2cm}
	\end{figure} 

\section{Experiment}
To generate high-resolution datasets for different problems investigated in this work, the Newton-Krylov-Schwarz algorithm in which the N-S Eq.~\ref{eq:ns1}-~\ref{eq:ns2} are approximated is employed to obtain the solution of the Jacobian system \cite{lin2021Ahighly}. Two kinds of dataset, including  synthetic and real-world, were generated to validate the performance of the proposed approach. For synthetic dataset, we consider the prototypical problem of a two dimensional flow past a circular cylinder. For real-world problems, we consider three dimensional blood flow in realistic hepatic portal vein and brain artery containing complex geometries. We compare our model with state-of-the-art PINNs, including Vanilla PINN \cite{rassi2019PINN}, NSFnets\cite{jin2021NSFnets} and PIPN\cite{kashefi2022PointNet} on above datasets. For NSFnets, we followed the vorticity-velocity formulation without pressure field output and the original hyperparameters. Since PIPN is designed specifically for two-dimensional problem, we only compare it with the proposed Pi-fusion on synthetic dataset. We train our model for 20,000 iterations using the Adam optimizer. All experiments are conducted on an NVIDIA GeForce RTX 3090. 

 \begin{figure}[t]
		\setlength{\belowcaptionskip}{-0.2cm}
		\centering
		\includegraphics[width=\linewidth]{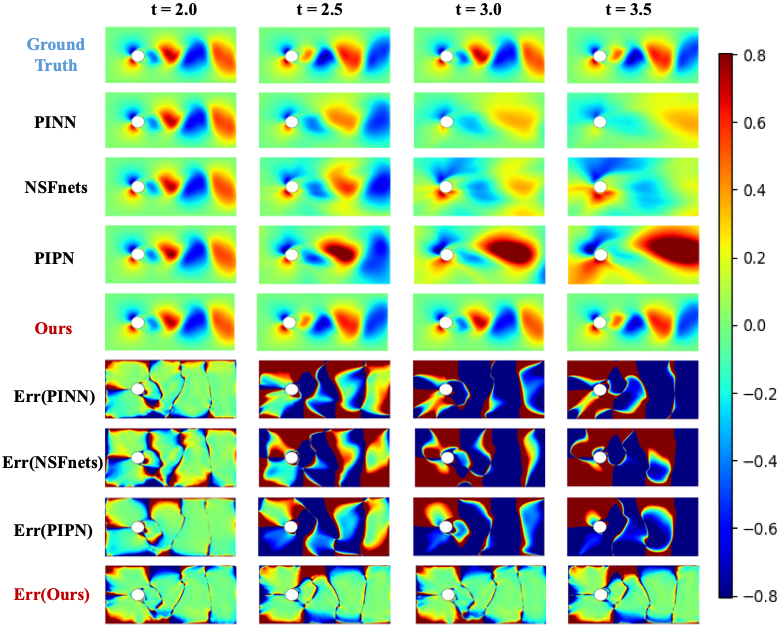}
            \vspace{-10pt}
		\caption{Comparison of the accuracy between Pi-fusion and other methods for the synthetic data in terms of y-direction velocity. Four representative time snapshots are chosen as the example ($t = 2.0s, 2.5s, 3.0s, 3.5s$). Err refers to the difference in the entire domain between ground truth (using direct numerical simulation) and prediction by the approach.}
		\label{result_v}
		\vspace{-0.2cm}
	\end{figure} 
 
 \begin{figure}[t]
		\setlength{\belowcaptionskip}{-0.2cm}
		\centering
		\includegraphics[width=\linewidth]{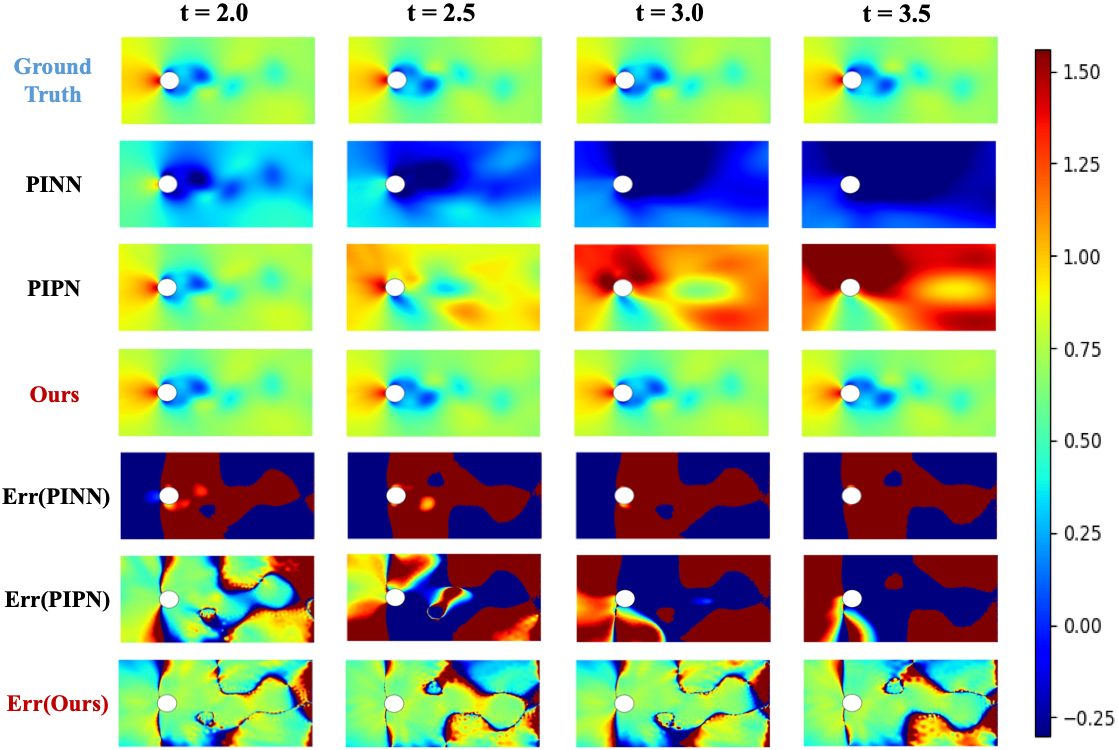}
            \vspace{-10pt}
		\caption{Comparison of the accuracy between Pi-fusion and other methods for the synthetic data in terms of pressure. Four representative time snapshots are chosen as the example ($t = 2.0s, 2.5s, 3.0s, 3.5s$). Err refers to the difference in the entire domain between ground truth (using direct numerical simulation) and prediction by the approach.}
		\label{result_p}
		\vspace{-10pt}
	\end{figure} 
 
 \begin{figure}[t]
		\setlength{\belowcaptionskip}{-0.2cm}
		\centering
		\includegraphics[width=\linewidth]{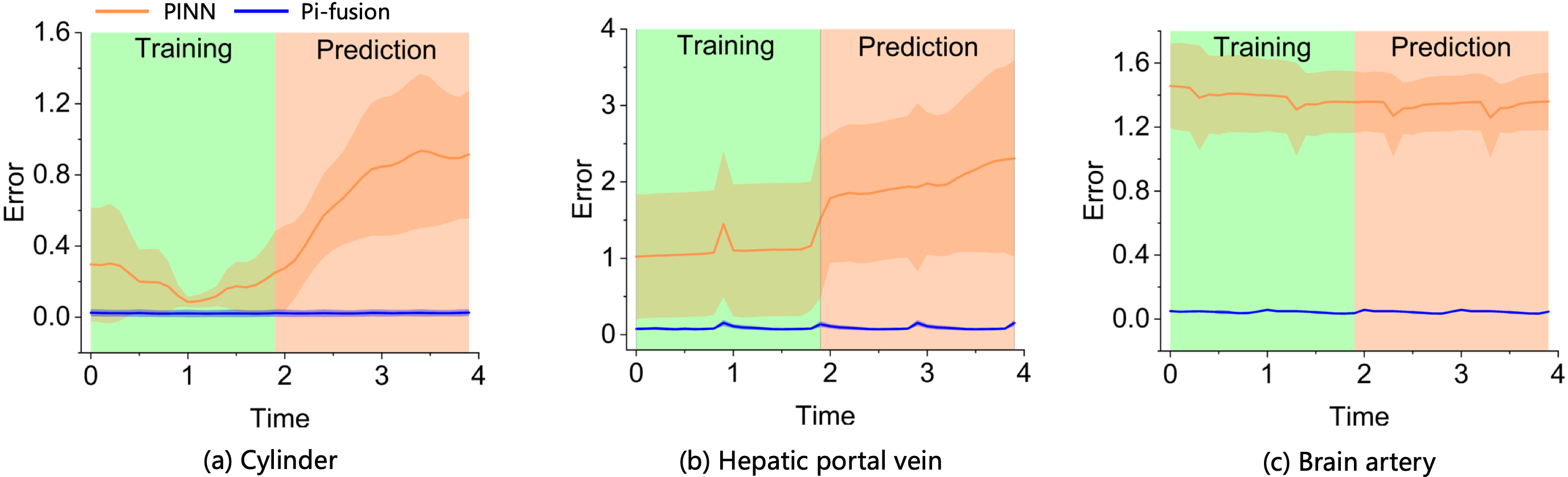}
        \vspace{-15pt}
		\caption{Relative prediction error of Pi-fusion and baseline PINN for learning fluid dynamics. The green and pink boxes correspond to the results included in the training and testing dataset. The shaded area shows the standard deviation of the relative prediction error.}
		\label{compare}
		\vspace{-0.05cm}
	\end{figure}
Five quantitative metrics are used to examine the performance of our method as presented in \cite{takamoto2022PDEBEACH}: (1) root-mean-squared-error($\textbf{RMSE}$), (2) $\textbf{normalized RMSE (nRMSE)}$, defined as $nRMSE = \lVert u_{pred} - u_{true}\rVert_{2} /\ \lVert u_{true}\rVert_{2}$ measures the $L_{2}$-norm distance between the prediction $u_{pred}$ and the ground truth $u_{true}$.  (3) $\textbf{maximum error}$, measures the model's worst prediction. While the aforementioned metrics assess the model's global performance, additional metrics are introduced to gauge specific failure modes: (4) $\textbf{RMSE of conserved value (cRMSE)}$ is defined to measure the error at physically conserved value. (5)$\textbf{RMSE at boundaries (bRMSE)}$ measures the error at the boundary, indicating if the model understand the boundary condition properly.

\begin{table*}[t]
	\centering
        \caption{Performance comparison between the proposed Pi-fusion and state-of-the-art PINNs}
        \renewcommand\arraystretch{1.5}
	\setlength{\abovecaptionskip}{0.2cm}
	\resizebox{\linewidth}{!}{
		\begin{tabular}{ccccccc}
			\hline
			Cases & Method & RMSE & nRMSE & max error & cRMSE & bRMSE\\
                \hline
			\multirow{6}{*}{2D Cylinder} & \textbf{Pi-fusion} & \boldmath{$9\times 10^{-3}$} & \boldmath{$2.23\times10^{-2}$} & \boldmath{$7.78\times10^{-2}$}& \boldmath{$1.07\times10^{-2}$} & \boldmath{$1.5\times10^{-3}$}\\ & PINN + reciprocal learning & $9.01\times10^{-2}$ & $2.18\times10^{-1}$ & $2.71\times10^{-1}$ & $2.18\times10^{-1}$ & $6.48\times10^{-3}$\\ & PINN\cite{rassi2019PINN} & $7.33\times10^{-1}$ & $4.08\times10^{-1}$ & $5.04\times10^{-1}$ & $4.1\times10^{-1}$ & $3.05\times10^{-1}$\\
   & NFSnets\cite{jin2021NSFnets} & $5.54\times10^{-1}$ & $2.25\times10^{-1}$ & $5.52\times10^{-2}$ & $1.78\times10^{-1}$ & $2.99\times10^{-2}$\\
   & PIPN\cite{kashefi2022PointNet} & $9.61\times10^{-1}$ & $7.42\times10^{-1}$ &$7.2\times10^{-1}$ & $6.66\times10^{-1}$ & $4.47\times10^{-1}$\\
                \hline
                \multirow{5}{*}{3D Hepatic portal vein } & \textbf{Pi-fusion} & \boldmath{$7.91
                \times10^{-2}$} & \boldmath{$5.51\times10^{-2}$} & \textbf{2.99} & \boldmath{$9.02\times10^{-2}$} & \boldmath{$2.48\times10^{-2}$}\\& PINN + reciprocal learning & $3.88\times10^{-1}$ & $7.07\times10^{-1}$ & 5.44 & 1.15 & $2.38\times10^{-1}$\\ & PINN \cite{rassi2019PINN}& $9.05\times10^{-1}$ & 1.49 & 7.18 & 2 & 1.21\\
                & NFSnets \cite{jin2021NSFnets}& $5.79\times10^{-1}$ & $8.58\times10^{-1}$ & 5.13 & 1.34 & $3.52\times10^{-1}$\\
                & PIPN \cite{kashefi2022PointNet} & — & —& — & — & —
                \\ 
                \hline
                \multirow{5}{*}{3D Brain artery} & \textbf{Pi-fusion} & \boldmath{$6.51\times10^{-2}$} & \boldmath{$3.18\times10^{-2}$} & \textbf{1.62} & \boldmath{$4.5\times10^{-2}$} & \boldmath{$2.33\times10^{-2}$}\\ 
                & PINN + reciprocal learning & 1.78 & 1.69 & 10.53 & 1.36 & 2.53\\ 
                & PINN \cite{rassi2019PINN}& 1.32 & 1.73 & 10.26 & 1.04 & 1.24\\
                & NFSnets \cite{jin2021NSFnets}& $8.79\times10^{-1}$ & 1.04 & 6.39 & 1.22 &$9.93\times10^{-1}$\\
                & PIPN \cite{kashefi2022PointNet} & — & — & — & — & —\\
			\hline
		\end{tabular}}
    \vspace{-10pt}
%
\label{performance}
\end{table*}

\begin{figure}[t]
\begin{minipage}[a]{1\linewidth}
    \centering
    \subfloat[]{\label{figpi1}\includegraphics[width=0.9\linewidth]{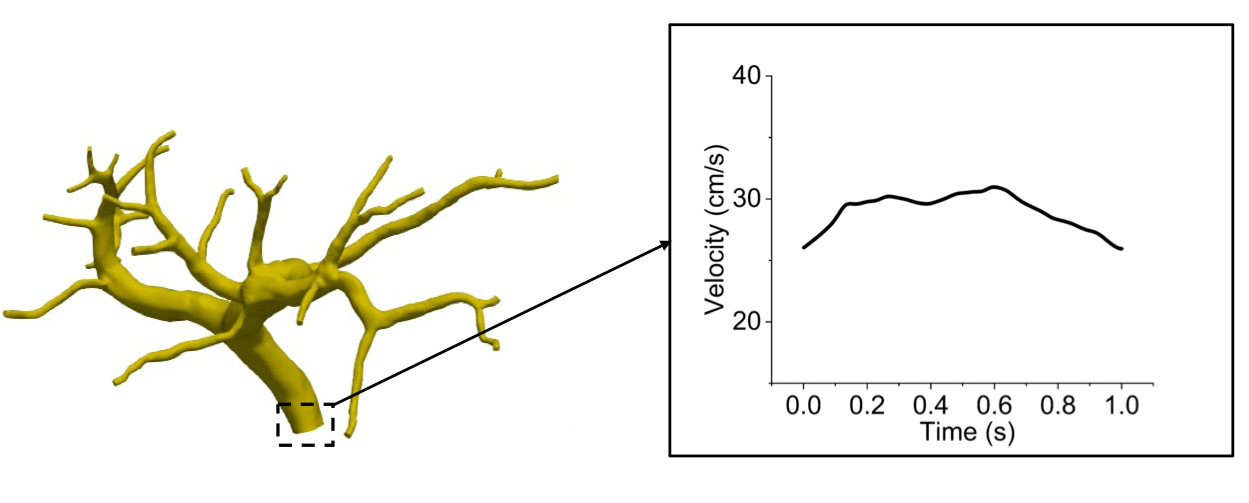}}
\end{minipage}
\begin{minipage}[b]{1\linewidth}
    \centering
    \subfloat[]{\label{figsa1}\includegraphics[width=0.9\linewidth]{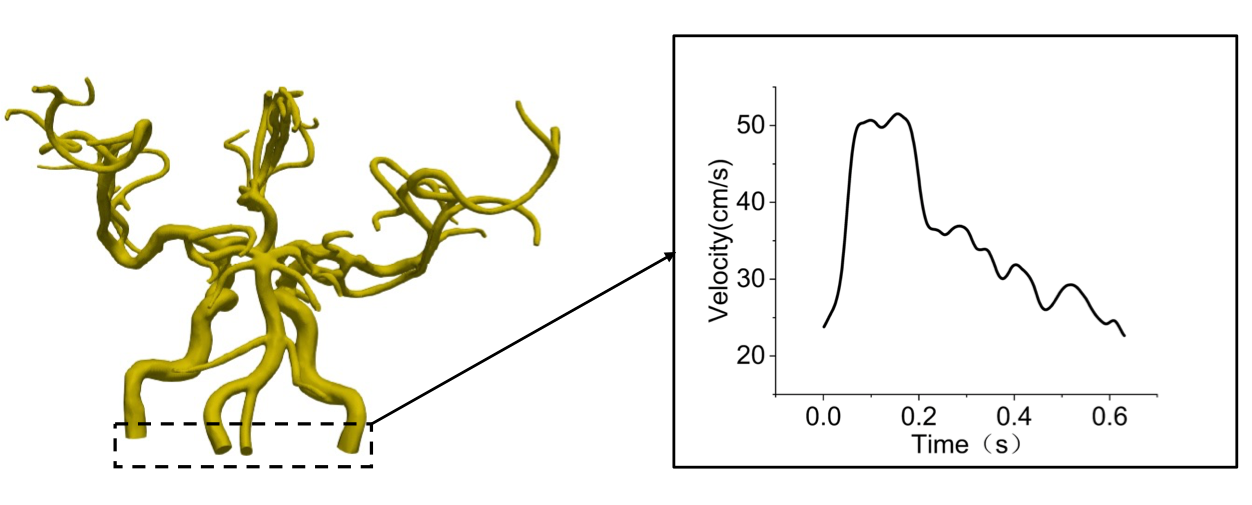}}
\end{minipage}
\vspace{-3pt}
\caption{3D patient-specific hepatic(a) and brain artery(b): The left panel shows the simulation domain at a time instant. The right panel shows that the physiologic flow waveform used as the inflow boundary condition.}
\label{fig:inflow}
\vspace{-9pt}
\end{figure}

\begin{table}[t]
	\centering
        \caption{Comparison of computational complexity of the proposed Pi-fusion with baseline methods including a single forward runs of traditional numerical method, PINN, PINN+reciprocal learning and NSFnet exemplified with 3D brain artery. The unit used for the time is seconds.}
        \renewcommand\arraystretch{1.5}
	\setlength{\abovecaptionskip}{0.2cm}
	\resizebox{\linewidth}{!}{
		\begin{tabular}{ccc}
			\hline
                Model & Inference & Parameter
			\\
                 \hline
                \textbf{Pi-fusion} &\textbf{0.03} & \textbf{6.387M} \\
                \hline
			{PINN \cite{rassi2019PINN}} & 0.26 & 8.672M\\
                \hline
                {NFSnets \cite{jin2021NSFnets}}  &0.15 & 9.742M \\
                \hline
                {Traditional numerical method}  &90.07 & — \\
			\hline
		\end{tabular}}
   \vspace{-15pt}	
\label{runningtime}
\end{table}

\subsection{Synthetic data}
\textbf{Data generation.} As the first example, we consider the prototypical problem of a two dimensional flow past a circular cylinder, exhibiting rich dynamic behavior and transitions for different regimes of the Reynolds number $Re = U_{\infty}D \backslash \nu$. Assuming a non-dimensional free stream velocity $U_{\infty} = 1$, cylinder diameter $D = 0.2$, and kinematic viscosity $\nu = 0.002$, the system is characterized by an asymmetrical vortex shedding pattern in the wake of the cylinder, known as the K$\acute{a}$rm$\acute{a}$n vortex street \cite{whitham1969AITFD}. We finally obtain a synthetic dataset lasting 4 seconds (40 time instants) by using direct numerical simulation, with a time step $\Delta t=0.1s$. 20\% of the total available data ($N = 4094$ points) on every time instants between $[0s, 2s)$ were selected as the train dataset, while 10\%  of them were selected as the validation dataset. Data on every time instants between $[2s, 4s)$ were selected as the test dataset to verify the performance of the proposed Pi-fusion. 

 \begin{figure*}[t]
\begin{minipage}[b]{1\linewidth}
    \centering
    \subfloat[]{\label{liver_20}\includegraphics[width=0.52\linewidth]{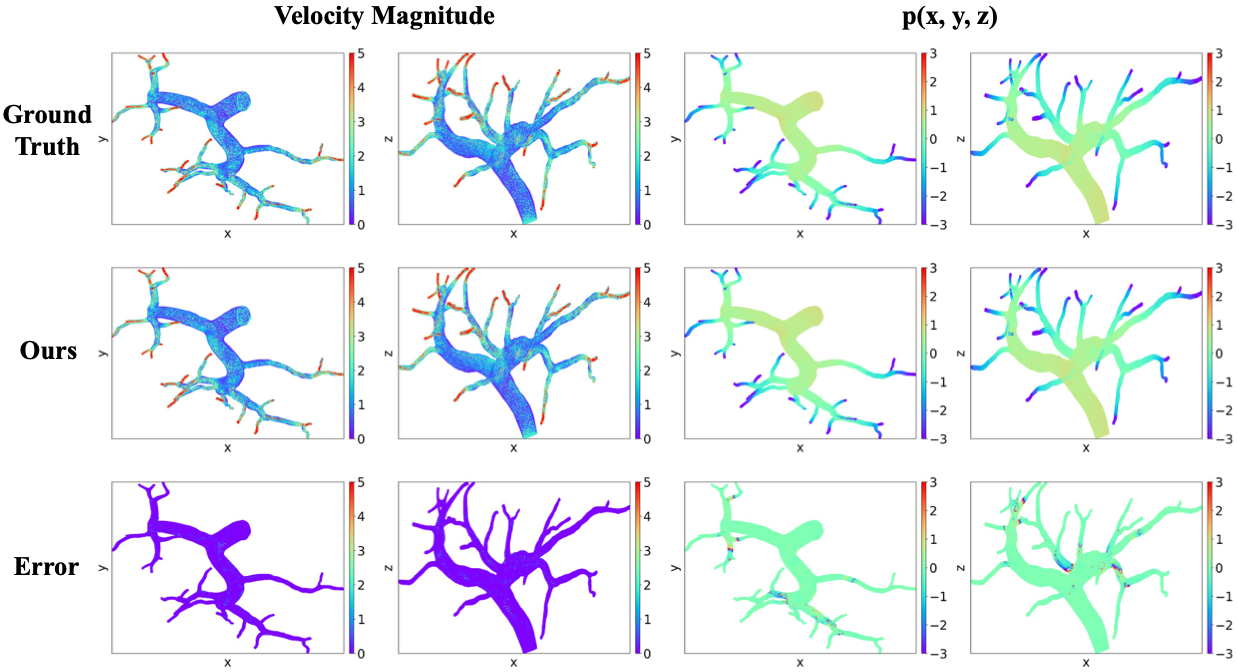}}
    \subfloat[]{\label{liver_25}\includegraphics[width=0.48\linewidth]{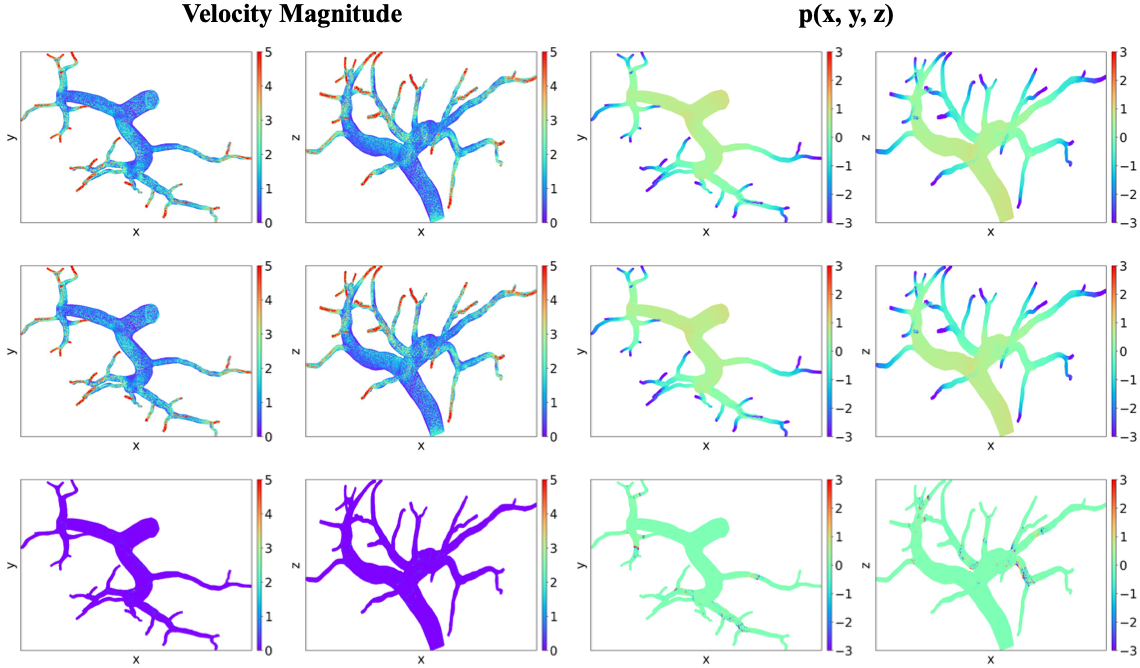}}
    \vspace{-10pt}
\end{minipage}
\begin{minipage}[b]{1\linewidth}
     \centering
    \subfloat[]{\label{liver_30}\includegraphics[width=0.52\linewidth]{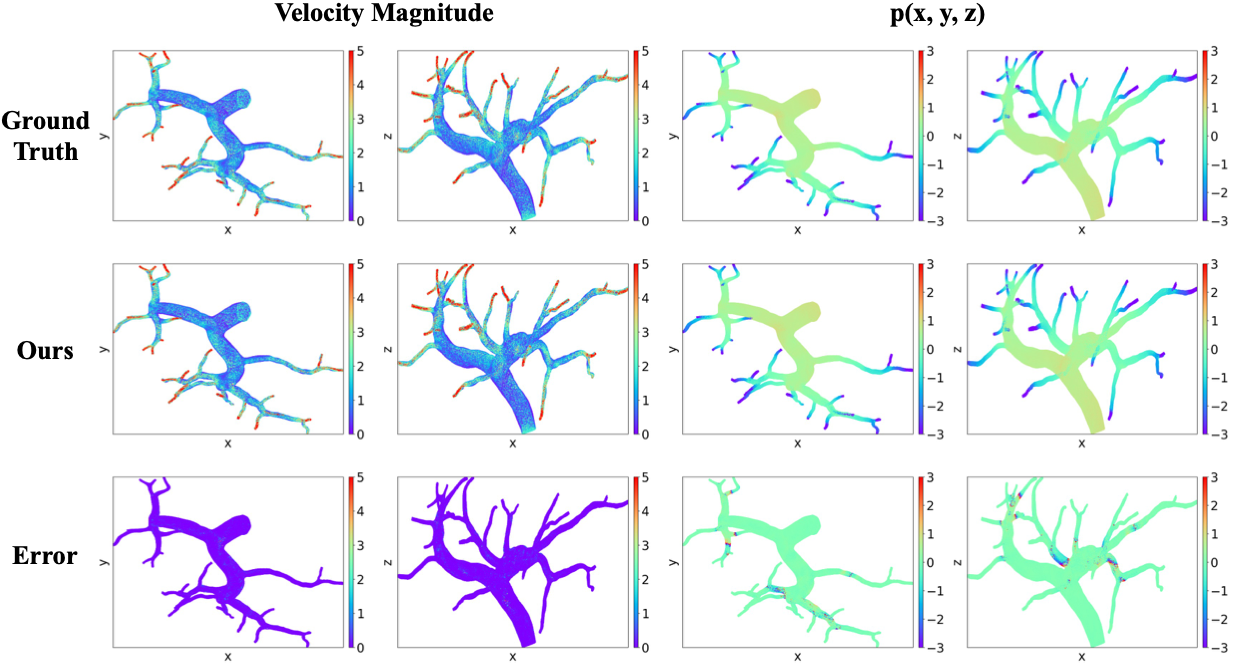}}
    \subfloat[]{\label{liver_35}\includegraphics[width=0.48\linewidth]{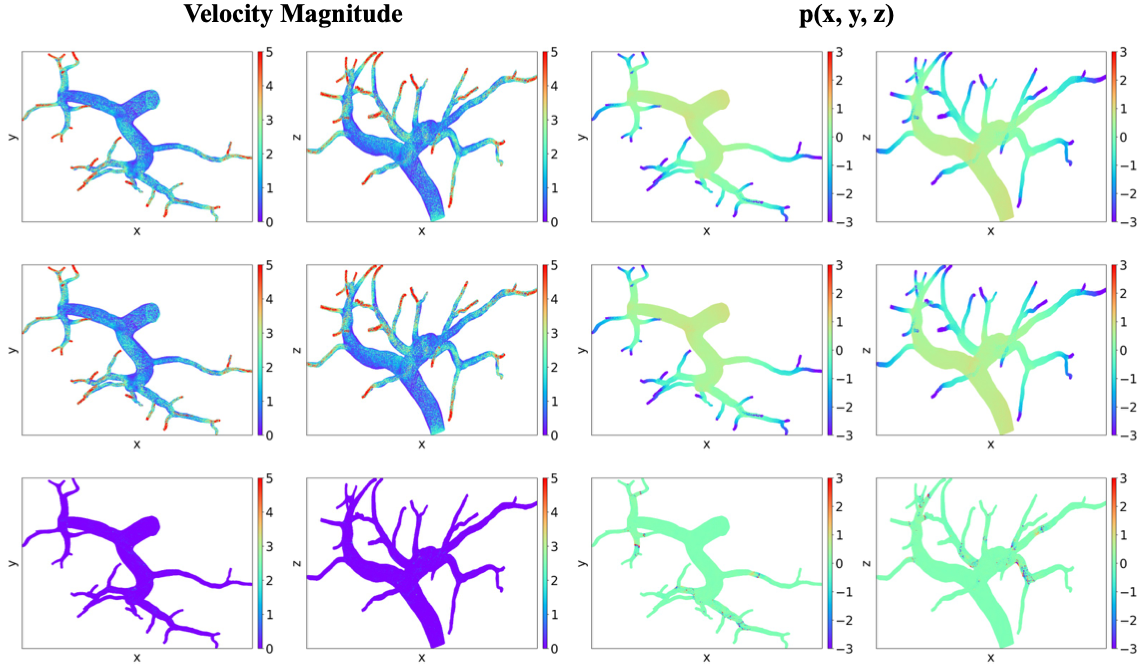}}
\end{minipage}
\vspace{-8pt}
\caption{3D hepatic portal vein: snapshots of $(a)t = 2.0s, (b)t = 2.5s, (c)t = 3.0s$ and $(d)t = 3.5s$ reference and predicted fields plotted on two perpendicular planes for velocity magnitude and pressure $p$.}
\label{fig:liver}
\end{figure*}
\textbf{Result.} 
As shown in Fig.~\ref{result_u}-Fig.~\ref{result_p}, good agreement can be achieved between the predictions of the proposed Pi-fusion and the reference, for typical time instants. From the estimation accuracy in terms of x-directed velocity presented in Fig.~\ref{result_u}, it can be observed obviously that the proposed approach exhibits better performance than other methods. It can be also observed that the prediction error at the boundary conditions and areas with significant variations is relatively large, indicating further research is needed. The relative prediction errors and their distributions over all time instants are presented in Fig.~\ref{compare}(a), it can be seen that compared with traditional PINN, the proposed Pi-fusion has significantly more robust performance (with steady relative error) over both the training and prediction stage. While with traditional PINN, the prediction is reasonable at time instant $t = 2.0s$ but degraded significantly in the following time instants. These results indicating strong generalization of the proposed Pi-fusion in predicting the temporal evolution of velocity and pressure fields. The quantitative results based on the evaluation metrics are summarized in Table~\ref{performance}. It can be seen that the proposed Pi-fusion significantly outperforms state-of-art PINNs, for both the entire spatial grids and the boundary conditions, confirming its good robustness. Specifically, the proposed Pi-fusion achieved smaller prediction error compared with state-of-the art PINNs (with a RMSE of 0.9\%, nRMSE of 2.23\%, max error of 7.78\%, cRMSE of 1.07\% and bRMSE of 0.15\%), indicating its remarkable performance for learning 2D fluid dynamics.
\vspace{-5pt}


\begin{figure*}[t]
\begin{minipage}[b]{1\linewidth}
    \centering
    \subfloat[]{\label{brain_20}\includegraphics[width=0.52\linewidth]{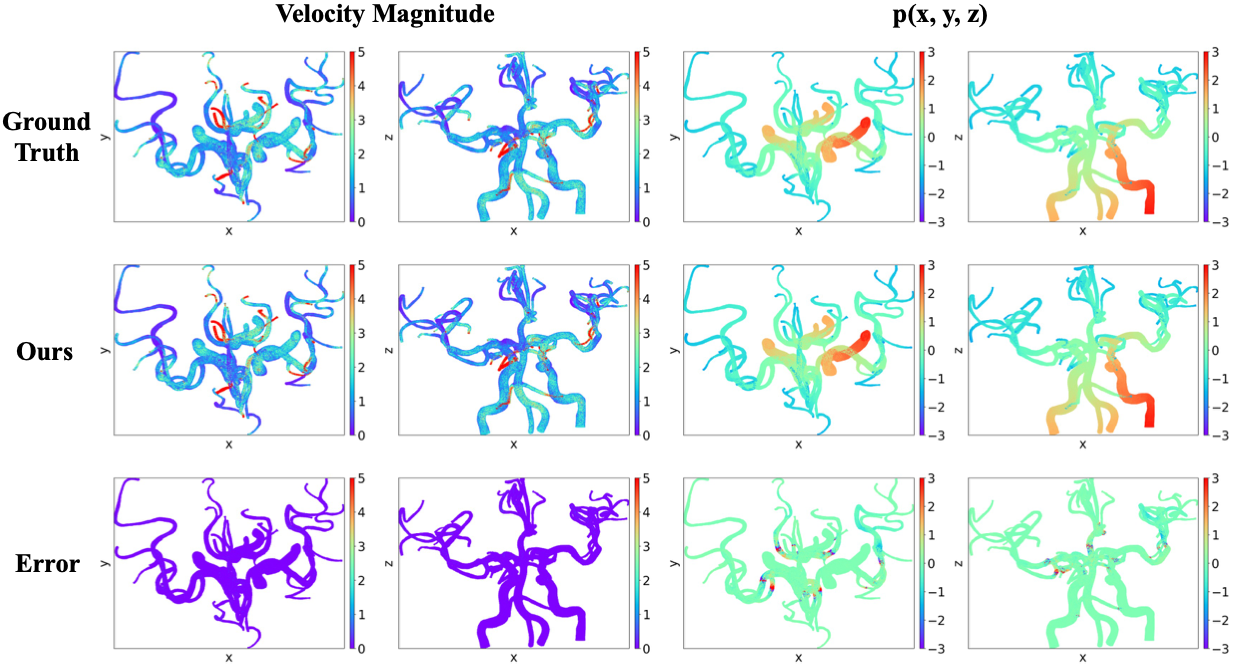}}
    \subfloat[]{\label{brain_25}\includegraphics[width=0.48\linewidth]{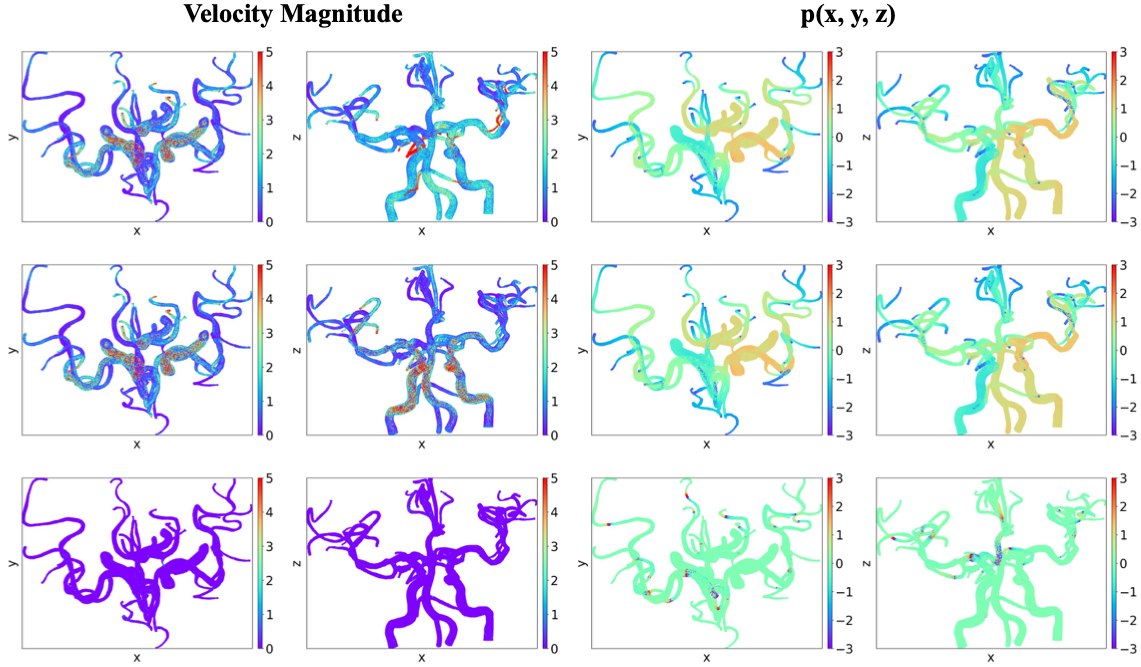}}
    \vspace{-10pt}
\end{minipage}
\begin{minipage}[b]{1\linewidth}
     \centering
    \subfloat[]{\label{brain_30}\includegraphics[width=0.52\linewidth]{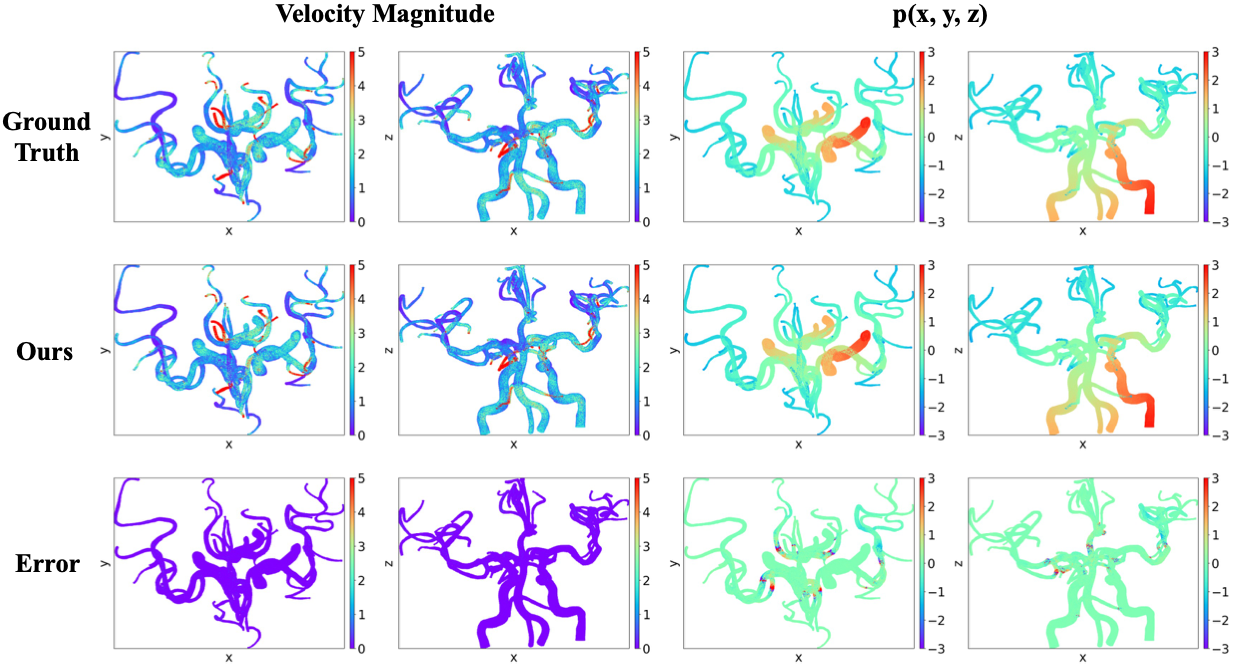}}
    \subfloat[]{\label{brain_35}\includegraphics[width=0.48\linewidth]{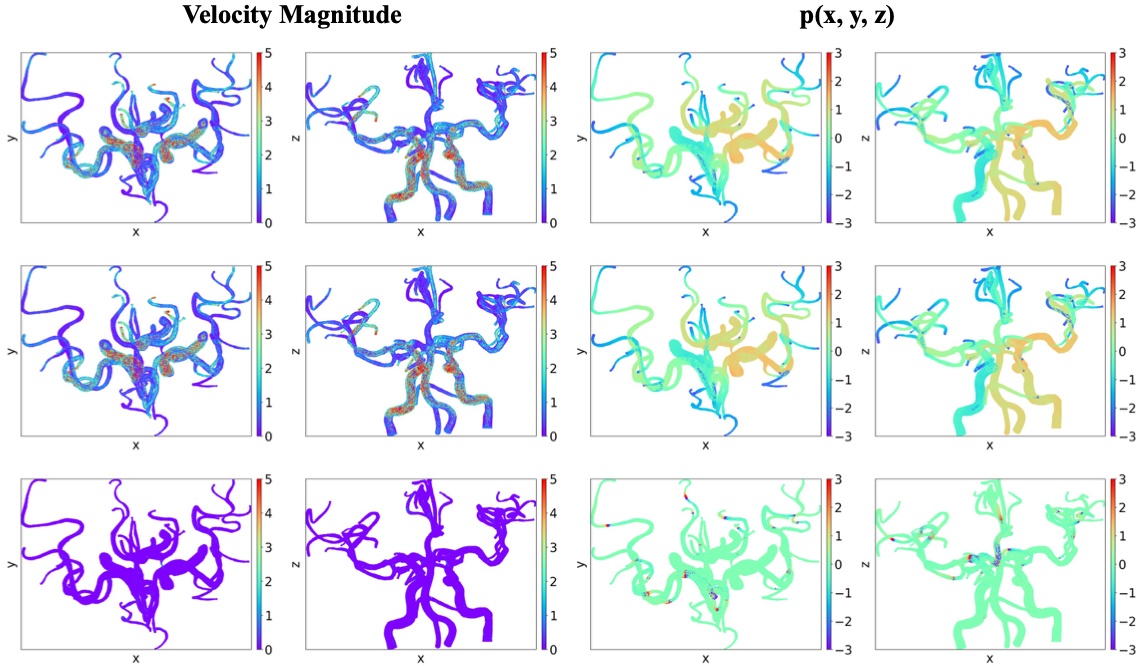}}
\end{minipage}
\vspace{-5pt}
\caption{3D brain artery: snapshots of $(a)t = 2.0s, (b)t = 2.5s, (c)t = 3.0s$ and $(d)t = 3.5s$ reference and predicted fields plotted on two perpendicular planes for velocity magnitude and pressure $p$.}
\label{fig:brain}
\end{figure*}

\subsection{Real-World data}
\vspace{-4pt}
\textbf{Data generation.} To further illustrate the performance of the proposed Pi-fusion in addressing real-world problems, three dimensional blood flow in realistic hepatic portal vein and brain artery were considered. Reference blood flow fields were generated using a fully implicit finite element method on an unstructured mesh \cite{lin2021Ahighly, lin2022Numerical}. The algorithm is implemented with the Portable Extensible Toolkit for Scientific computation library. For generating the flow filed in 3D Hepatic portal vein, $\rho = 1.05 g/cm^3$ and $\nu = 0.038 cm^2/s$ are used. The hepatic portal vein has 1 inlet and 42 outlets. For 3D Brain artery, $\rho$ is set to $1.06 g/cm^3$ and $\nu$ is set to $0.035 cm^2/s$. The simulation domain and the specific inflow velocity boundary condition of real-world data are shown in Fig.\ref{fig:inflow}, the total resistance $R = 100 dynes\cdot s/cm^5$ is chosen such that the computed pressures is within the ranges of typical values in adults. We finally obtain two real-world datasets, both lasting 4 seconds (40 time instants), with a time step $\Delta t=0.1s$. For both dataset, 10\% of the total available data on every time instants ($N = 507936$ points for 3D Hepatic portal vein and 540824 for 3D Brain artery) between $[0s, 2s)$ were selected as the train dataset, while 10\%  of them were selected as the validation dataset. Data on every time instants between $[2s, 4s)$ were selected as the test dataset to verify the performance of the proposed Pi-fusion in real-world scenarios.


\textbf{Results.} As shown in Fig.~\ref{fig:liver} and Fig.~\ref{fig:brain}, good agreements are achieved between the predictions on velocity and pressure field of the proposed Pi-fusion and the reference, for both 3D Hepatic portal vein and 3D Brain artery. Similar to the synthetic case, the proposed Pi-fusion shows significantly more robust performance for real-world datasets (Fig.~\ref{compare}(b) and Fig.~\ref{compare}(c)) than traditional PINN. We also observe that PINN and NFSnets fails to produce accurate predictions for real-world complex flow fields, even during the training stage. From Table ~\ref{performance} we can see, the proposed Pi-fusion outperforms the other two approaches with significantly lower prediction errors. The proposed Pi-fusion achieves smaller prediction errors for both 3D Hepatic portal vein  (with a RMSE of 7.91\%, nRMSE of 5.51\%, cRMSE of 9.02\% and bRMSE of 2.48\%) and 3D Brain artery (with a RMSE of 6.51\%, nRMSE of 3.18\%, cRMSE of 4.5\% and bRMSE of 2.33\%), highlighting its good performance for learning real-world fluid dynamics. 

We also present the detailed comparison of computational complexity between Pi-fusion and the baseline models used in this work, summarized in Table ~\ref{runningtime}, with 3D brain artery as the example. The same hardware resources were used in different deep learning based methods to calculate the running time for fair comparison. The inference time of Pi-fusion is only 0.03s, which is significantly lower than state-of-the-art PINNs. Furthermore, compared with traditional numerical method, the computational efficiency was improved by three orders of magnitude with the proposed Pi-fusion.

\begin{table*}[t]
	\begin{center}
 \caption{Ablation studies with 2D synthetic data as the example for different settings.}
	\setlength{\abovecaptionskip}{0.2cm}
        \resizebox{\linewidth}{!}{
		\begin{tabular}{c|c|c|c|c|c|c|c|c}
			\hline
			Setting & Diffusion & Reciprocal & Way to use PDE & RMSE & nRMSE & max error & cRMSE & bRMSE\\
			\hline 
            1 & \CheckmarkBold& \CheckmarkBold & PDE guidance& $9\times 10^{-3}$ & $2.23\times10^{-2}$ & $7.78\times10^{-2}$ & $1.07\times10^{-2}$ & $1.5\times10^{-3}$\\
		   2 & \XSolidBrush& \CheckmarkBold & PDE guidance& $1.04\times10^{-2}$ & $2.58\times10^{-2}$ & $8\times10^{-2}$ & $2.57\times10^{-2}$ & $5.93\times10^{-2}$\\ 
             3 & \CheckmarkBold & \XSolidBrush & PDE guidance&$2.39\times10^{-1}$ & $9.27\times10^{-1}$ & $5.93\times10^{-1}$ & $9.33\times10^{-1}$ & $4.26\times10^{-1}$\\
              4 & \CheckmarkBold & \CheckmarkBold & PDE loss & $3.01\times10^{-1}$ & $9.24\times10^{-2}$ & $2.57\times10^{-1}$ & $9.25\times10^{-2}$ & $1.48\times10^{-1}$\\ 
	       5 & \CheckmarkBold & \CheckmarkBold & \XSolidBrush& $1. 09\times10^{-2}$ & $2.61\times10^{-2}$ & $8.08\times10^{-2}$ & $2.88\times10^{-2}$ & $5.93\times10^{-2}$\\ 
        
			\bottomrule 
		\end{tabular}
        }
	\label{abstudy}
	\end{center}
 \vspace{-20pt}
\end{table*}

\label{gen_inst}

\subsection{Ablation studies}
To investigate the contribution of the proposed components to the overall performance, we conduct the ablation studies by taking 2D synthetic data as the example (Table ~\ref{abstudy}). Setting 2: we compare our model with its counterpart without diffusion process. The RMSE of the model without diffusion process is increased by 15\%, showcasing the advantages of diffusion process in distribution fitting ability. Setting 3: The performance of the approach without the training strategy degraded significantly, indicating its better generalization by introducing the reciprocal learning component. Setting 4\&5: We examine the importance of the physics-informed guidance sampling method by introducing two settings, removing the PDE term and replacing it with PDE loss as traditional PINN. Without the guidance of physical mechanism, both models might fail to capture the temporal dynamics and thus show significant degraded performance (RMSE of 30\% for setting 4 and 1.09\% for setting 5). 

\vspace{-10pt}

\section{Conclusion}
\textbf{Summary} In this work, we proposed a physics-informed diffusion model, Pi-fusion, to learn fluid dynamics. The physics-informed guidance together with diffusion process bring the Pi-fusion significant advantages in generalizability in predicting temporal fluid motion, and interpretability. The performance have been demonstrated over both synthetic and real-world fluid dynamics on incompressible Newtonian flows, compared to the state-of-the-art PINNs. Experimental results verified that the proposed Pi-fusion has promising potential to serve as a reliable surrogate model for learning imcompressible Newtonian dynamics. Our work will also open a new era of the utilization of diffusion models for learning physical dynamics.

\textbf{Limitations and future work} Even though the proposed approach showed remarkable performance in learning fluid dynamics, there are still several limitations. First, due to the huge cost in generating the reference fluid fields, the proposed approach is only validated in the consecutive 4 seconds, whether the performance remain stable after a long time should be further studied. Second, although the proposed approach achieved small average prediction error over the entire geometry in real-world dataset, we found that the prediction errors at the vascular branch is still relatively large. The possible reason is that the flow field at the branch involves more complicated situations such as inconsistent velocity directions at adjacent points, indicating dedicated optimization methods should be designed.

{\small
		\bibliographystyle{unsrt}
		\bibliography{neurips_2022}
	}

\clearpage
\end{document}